# Selectively Linearized Neural Network based RoCoF-Constrained Unit Commitment in Low-Inertia Power Systems

Mingjian Tuo
*Student Member, IEEE*
Department of Electrical and Computer Engineering
University of Houston
Houston, TX, USA
mtuo@uh.edu

Xingpeng Li
*Senior Member, IEEE*
Department of Electrical and Computer Engineering
University of Houston
Houston, TX, USA
xli82@uh.edu

*Abstract*—Conventional synchronous generators are gradually being replaced by inverter-based resources, such transition introduces more complicated operation conditions. And the reduction in system inertia imposes challenges for system operators on maintaining system rate-of-change-of-frequency (RoCoF) security. This paper presents a selectively linearized neural network (SNLNN) based RoCoF-constrained unit commitment (SLNN-RCUC) model. A RoCoF predictor is first trained to predict the system wide highest locational RoCoF based on a high-fidelity simulation dataset. Instead of incorporating the complete neural network into unit commitment, a ReLU linearization method is implemented on active selected neurons to improve the algorithm computational efficiency. The effectiveness of proposed SLNN-RCUC model is demonstrated on the IEEE 24-bus system by conducting time domain simulation on PSS/E.

## I. INTRODUCTION

The decarbonization of the electricity generation relies on the integration of converter-based renewable energy sources (RES) over the past decades. In addition, the development of high-voltage direct current (HVDC) transmission systems has resulted in deployment of power-electronic devices in modern power system [1]. Current power systems may ultimately shift towards power systems where all generation comes from converter-based resources. And in sufficient inertia of the system has been found to be the main challenge of power systems stability [2].

Due to the retirement and replacement of conventional generation, more generation is coming from converter-based resources. Consequently, the system kinetic energy traditionally contributed by rotating rotor would decrease significantly, leaving the system more vulnerable to high rate of change of frequency (RoCoF) and large frequency excursion when a power disturbance occurs. When RoCoF violates the pre-specified threshold, protection devices would disconnect generators from the grid [3]. Fast frequency response services have been used to ensure frequency stability of system with reduced inertia [4]. Studies in [5] also show that virtual inertia and altering RoCoF protection are very effective in reducing the frequency stability risk of future converter-based power system in Ireland. Recently, transmission system operators (TSO) have imposed extra RoCoF related constraints in the conventional unit commitment model to keep the minimum amount of synchronous inertia online [6]. EirGrid has also introduced a synchronous inertial response constraint to ensure that the available inertia is above a minimum limit of 23 GWs in Ireland [7]. The Swedish TSO once ordered one of its nuclear power plants to reduce output by 100 MW to mitigate the risk of loss of that power plant [8]. Ref. [9] implemented a system equivalent model-based enhanced frequency stability constrained multiperiod optimal power flow model. In [10], frequency-related constraints were incorporated into security constrained unit commitment (SCUC) enforcing limitations on RoCoF that is derived from a uniform frequency response model. Ref. [11] studied a mixed analytical-numerical approach based on multi-regions and investigated a model combining evolution of the center of inertia and certain inter-area oscillations. Despite the great achievements in system frequency stability, these works only focus on the uniform equivalent model of the system dynamic which fails to incorporate the nodal inertial response and the impact of disturbance propagation [12]. To handle higher order characteristics of system dynamics, a deep neural network (DNN) based frequency-constrained unit commitment (FCUC) is proposed in [13] which incorporates frequency related constraints against the worst-case contingency. Ref. [14] proposed a DNN based RoCoF constrained unit commitment (DNN-RCUC) to handle the locational stability of the system. However, the computational efficiency of DNN-RCUC has not been discussed in previous work.

To bridge the aforementioned gaps, selectively linearized neural network based RoCoF-constrained unit commitment (SLNN-RCUC) model is first proposed, in which the derived frequency related constraints are incorporated into SCUC to secure the system locational frequency stability against worst $G-1$ contingencies. Secondly, a model-based data generation approach is used to efficiently generate practical cases for RoCoF predictor training while avoid divergency in time domain simulations. The generated dataset covers vast ranges of reasonable operating conditions. Thirdly, we propose an active neuron selecting method to reduce the approximation error of

ReLU linearization, such method improves the computational efficiency of the DNN-RCUC model while maintaining the model's high performance.

The remainder of this paper is organized as follows. Section II discusses the power system mathematical based model and RoCoF predictor model thoroughly. Section III presents the methodology of model-based data generation, ReLU linearization approaches and the proposed SLNN-RCUC model. Section IV shows the simulation results. Section V concludes this paper and presents future work.

## II. SYSTEM FREQUENCY DYNAMICS

### A. System Equivalent Model

The frequency of the power system is one of the most important metrics that indicate the system stability. Traditionally, the frequency is treated as unique of the whole system, which is derived from the system equivalent model extended from one-machine swing equation. The kinetic energy $E_i$ stored in a rotational mass is proportional to its moment of inertia and the square of its angular velocity:

$$E_i = \frac{1}{2} J_i \omega_i^2 \tag{1}$$

where $J_i$ is the moment of inertia of the shaft in kg·m²s; and $\omega_i$ is the nominal rotational speed. System inertia is defined as the total amount of rotational energy stored in all connected synchronous units,

$$E_{sys} = \sum_{i=1}^{N} \frac{1}{2} J_i \omega_i^2 = \sum_{i=1}^{N} H_i S_{B_i} \tag{2}$$

where $H_i$ is the inertia constant of the generator in seconds; $S_{B_i}$ is the base power in MVA;

The generator dynamics are described by the swing equation as:

$$P_m - P_e = M_i \frac{d\Delta\omega_i}{dt} + D_i \Delta\omega_i \tag{3}$$

where $P_m$ is the mechanical power and $P_e$ is the electrical output power, with $M_i$ and $D_i$ denoting the normalized inertia and damping constants. Uniform frequency response model has been widely used for system operations and the simplified swing equation of the whole grid is considered as the extension of one machine model [15].

### B. System Dynamic Model

Recent work has realized the danger of only considering the dynamics of the equivalent model in systems, which would underestimate the actual need for frequency ancillary services, leading to higher locational RoCoF and larger regional frequency deviation than expected.

Using the topological information and the system parameters, the transmission network can be modeled as a graph consisting of nodes (buses) and edges (branches). The oscillatory behavior of each individual bus can be expressed as follows,

$$m_i \ddot{\theta}_i + d_i \dot{\theta}_i = P_{in,i} - \sum_{j=1}^{n} b_{ij} \sin(\theta_i - \theta_j), \quad i \in \{1, \dots, n\} \tag{4}$$

where $m_i$ and $d_i$ denote the inertia coefficient and damping ratio for node $i$ respectively, while $P_{in,i}$ denotes the power input, and $b_{ij}$ denotes the line susceptance. With inertia on certain nodes $m_i > 0$, it is an approximation model for the swing dynamics of high-voltage transmission network within a short period following the event [16]. A network-reduced model with $N$ generator buses can be obtained by eliminating passive load buses via Kron reduction [17]. The phase angle $\theta$ of generator buses can be expressed by the following dynamic equation,

$$M \ddot{\theta} + D \dot{\theta} = P - L \theta \tag{5}$$

where $M = \text{diag}(\{m_i\})$, $D = \text{diag}(\{d_i\})$; for the Laplacian matrix $L$, its off-diagonal elements are $l_{ij} = -b_{ij} V_i^{(0)} V_j^{(0)}$, and diagonals are $l_{ij} = \sum_{j=1, j \neq i}^{n} b_{ij} V_i^{(0)} V_j^{(0)}$. Under the assumption of homogeneous inertia, the RoCoF within $\Delta t$ at bus $i$ can then be derived [16],

$$R_i(t) = \frac{\Delta P e^{-\frac{\gamma t}{2}}}{2\pi m} \sum_{\alpha=1}^{N_g} \frac{\beta_{\alpha i} \beta_{\alpha b}}{\sqrt{\frac{\lambda_\alpha}{m} - \frac{\gamma^2}{4}} \Delta t} \left[ e^{-\frac{\gamma \Delta t}{2}} \sin\left(\sqrt{\frac{\lambda_\alpha}{m} - \frac{\gamma^2}{4}} (t + \Delta t)\right) - \sin\left(\sqrt{\frac{\lambda_\alpha}{m} - \frac{\gamma^2}{4}} t\right) \right] \tag{6}$$

where $\lambda_\alpha$ is the eigenvalue of matrix $L$, $\beta_{\alpha i}$ is the eigen vector value; and $m$ denotes average inertia distribution on generator buses; while bus $b$ is where disturbance occurs. The ratio of damping coefficient to inertia coefficient $\gamma = d_i/m_i$ is assumed as a constant.

### C. RoCoF Prediction

Considering a system with $N_G$ generators, following a generator contingency $\varpi_t$ at period $t$, the highest locational RoCoF value of the system $R_t$ is considered as a nonlinear function with respect to the contingency level, contingency location, system states, unit dispatch, load data and RES profile [13].

$$R_t = h^{rcf}(s_t, u_t, d_t, r_t, \varpi_t) \tag{7}$$

where $s_t$ denotes the system states; $d_t$ is the load profile and $r_t$ is the RES forecast; and $u_t$ is the generation dispatch at period $t$. Now we use a DNN based RoCoF predictor for $h^r$ to replace the original nonlinear RoCoF function to constrain system wide RoCoF value, which is then expressed as,

$$\hat{R}_t = \hat{h}^{rcf}(x_t, W, b) \leq -RoCoF_{lim} \tag{8}$$

where $x_t$ is the feature vector of the system; $W$ and $b$ denote the well-trained parameters. $RoCoF_{lim}$ is the predefined threshold. We first consider that the inertial response of the system is determined by the generator pre-contingency status. Therefore, the generator status feature vector for case at period $t$ is first defined as follows,

$$u_t = [u_{1,t}, \cdots, u_{g,t}, \cdots, u_{N_G,t}] \tag{9}$$

Equation (6) shows that both the magnitude and location of the contingency will have impact on the RoCoF values. The loss of largest generation not only causes mismatch in system power balance but also degrades the system synchronous inertia, resulting in higher frequency deviation and larger initial RoCoF. Thus, the disturbance feature vector is defined against the loss of then largest generation, the magnitude of the contingency is expressed as,

$$P_t^\varpi = \max_{g \in G}(P_{1,t}, \cdots, P_{g,t}, \cdots, P_{N_G,t}) \quad (10)$$

The location of the disturbance is represented by the index of the generator producing maximum power,

$$g_t^\varpi = \arg\max_{g \in G}(P_{1,t}, \cdots, P_{g,t}, \cdots, P_{N_G,t}) \quad (11)$$

The disturbance feature vector is then defined as,

$$\varpi_t^G = [0, \cdots, 0, \underbrace{P_t^\varpi}_{g_t^\varpi th\ element}, 0, \cdots, 0] \quad (12)$$

Laplacian matrix $L$ of the grid and Fiedler mode value depend on the power-angle characteristics, which are determined by the active power injection [18]. Thus, the active power injection of all synchronous generators is also encoded into the feature vector.

$$P_t = [P_{1,t}, \cdots, P_{2,t}, \cdots, P_{N_G,t}] \quad (13)$$

The overall feature vector of a sample $s$ can be then defined as follows,

$$x_t = [u_t, \varpi_t^G, P_t] \quad (14)$$

### III. METHODOLOGY

#### A. Model-based Data Generation

Randomly power injections are utilized in [13] to ensure reliability under vast ranges of operating conditions. However, such method may subject to instability conditions. And divergency during the simulation initialization process would be another concern for dataset generation. To handle that issue, a model-based systematic data generation approach is used to generate reasonable and representative data that will be used to train RoCoF Predictors.

Training samples are generated from three models mentioned in [18] over various load and RES profiles: (1) Traditional SCUC (T-SCUC) is the base model without frequency related constraints; (2) System equivalent model based RoCoF constrained SCUC (ERC-SCUC) secure system aggregated frequency stability against $G - 1$ event; (3) The location based RoCoF constrained SCUC (LRC-SCUC) introduces locational RoCoF constraints based on the definition of local buses and non-local buses. Nodal RoCoF constraints ensure system stability by imposing limit on locational RoCoF over all buses under all $G - 1$ contingency. The details of all models are presented in [18].

#### B. DNN-RCUC Formulations

The objective of DNN-RCUC is to minimize the total system cost consisting of variable fuel costs, no-load costs, start-up costs, and reserve costs.

$$\min_\Phi \sum_{g \in G} \sum_{t \in T} (c_g P_{g,t} + c_g^{NL} u_{g,t} + c_g^{SU} v_{g,t} + c_g^{RE} r_{g,t}) \quad (15)$$

Besides constraints on generators and branches [18], we also encode the DNN into the MILP SCUC problem, decision variables are introduced to build the disturbance feature vector. binary variable $\eta_{g,t}^G$ is used to indicate the status of largest output power of generator $g$ in scheduling period $t$, a big-M method is introduced to express the disturbance vector $[\varepsilon_{1,t}, \cdots, \varepsilon_{N_G,t}]$ [14]. Thus, the input feature vector can be expressed as follows,

$$x_t = [u_{1,t}, \cdots, u_{N_G,t}, \varepsilon_{1,t}, \cdots, \varepsilon_{N_G,t}, P_{1,t}, \cdots, P_{N_G,t}] \quad (16)$$

RoCoF-limiting constraints can be derived from the pre-trained RoCoF predictor $\hat{h}^r(x,W,b)$. A fully connected neural network with $N_L$ hidden layer is considered with the following setting: each layer uses a ReLU activation function as $\sigma(\cdot) = \max(\cdot, 0)$, which returns zero when the node is with a negative value, implying the node is inactive. $x_t$ is fed into the well-trained predictor, and the output layer of the predictor is a linear activation function. the nonlinear constraints can be incorporated into MILP problems by introducing auxiliary binary variables $a$. The reformulation of DNN-based RoCoF predictor is expressed as follows,

$$\hat{z}_{1[l],t} = x_t W_{1[l]} + b_{1[l]}, \forall l, t, \quad (17a)$$

$$\hat{z}_{q[l],t} = z_{q-1,t} W_{q[l]} + b_{q[l]}, \forall l, t, \quad (17b)$$

$$z_{q[l],t} \leq \hat{z}_{q[l],t} - M(1 - a_{q[l],t}), \forall q, l, t, \quad (17c)$$

$$z_{q[l],t} \geq \hat{z}_{q[l],t}, \forall q, \forall l, \forall t, \quad (17d)$$

$$z_{q[l],t} \leq M a_{q[l],t}, \forall q, \forall l, \forall t, \quad (17e)$$

$$z_{q[l],t} \geq 0, \forall q, \forall l, \forall t, \quad (17f)$$

$$a_{q[l],t} \in \{0,1\}, \forall q, \forall l, \forall t, \quad (17g)$$

$$\hat{R}_t = z_{L,t} W_{L+1} + b_{L+1}, \forall t, \quad (17h)$$

where $l$ indicates the index of neurons, and $q$ represents the network layer; $z$ and $\hat{z}$ represent the activated value and pre-activated value of each neuron respectively. The RoCoF related constraint considering threshold $-\text{RoCoF}_{\text{lim}}$ can be then formulated as,

$$\hat{R}_t \leq -RoCoF_{lim}, \forall t, \quad (17i)$$

#### C. Active ReLU Linearization

In a deep neural network, there is a set of layers of neurons; activation function, such as ReLU, is applied to the result of linear combination of values from neuron nodes [19]. Previous study in [14] has found that incorporation of DNN would introduces multiple binary variables, as explained in section III.B. Such reformulation increases the computational burden of the SCUC model.

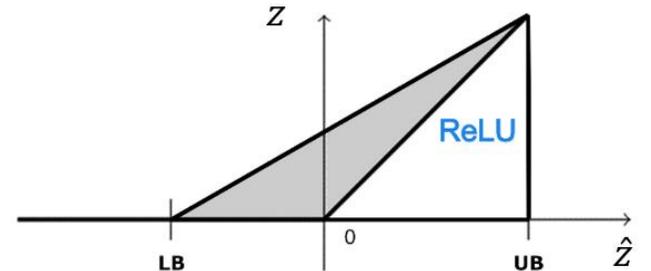

Figure. 1. The activation function of a ReLU node and its approximation.

Reference [20] has demonstrated that linearization of ReLU function can greatly reduce the DNN size without too much degradation of classification accuracy. Approximation of ReLU function is shown in Fig. 1. The weighted sum of input feature to the node is denoted as variable $\hat{z}$, and the output of the node is denoted using the variable $z$. Given the upper and lower bounds $[LB, UB]$ of $\hat{z}$. The relationship of z and x can then be approximated by a set of constraints $z \geq 0$, $z \geq \hat{z}$, and $z \leq \frac{UB \cdot (\hat{z} - LB)}{UB - LB}$. These constraints are all linear equations with constant $UB$ and $LB$.

To reduce the approximation error, one mitigation is to add penalty on nonnegative $z$ in the objective function to push activated $z$ to be set at the bottom two sides of the triangle. Intuitively, ReLU linearization is applied to each neuron node, converting a DNN-RCUC model into ReLU linearized neural network based RCUC (RLNN-RCUC). However, this process may introduce approximation error and subsequently result in low prediction accuracy. In this work, we propose an active selecting method to apply linearization on selected neurons, improving the computational efficiency of the SCUC model while reducing the approximation error. The generated out of sample dataset is first fed into the well-trained network, a nodal positivity index $\varepsilon_{q[l]}$ is proposed to estimate the percentage of positive preactivated values of neuron node $l$ in $q$ layer.

$$\varepsilon_{q[l]} = \frac{1}{N}\left(\sum_N \hat{z}_{q[l],n} - \sum_N \left|\hat{z}_{q[l],s} - \frac{1}{N}\sum_N \hat{z}_{q[l],n}\right|\right) \quad (18)$$

A threshold of 0.5 is set to select nodes suitable for ReLU linearization with less approximation error. Thus, (17c)-(17g) for selected neurons in set $\mathcal{H}$ can be replaced by (19a)-(19c), which could be applied to convert DNN-RCUC model into a SLNN-RCUC model.

$$z_{q[l],t} \geq \hat{z}_{q[l],t}, \forall q, \forall l \in \mathcal{H}, \forall t, \quad (19a)$$

$$z_{q[l],t} \leq \frac{UB \cdot (\hat{z}_{q[l],t} - LB)}{UB - LB}, \forall q, \forall l \in \mathcal{H}, \forall t, \quad (19b)$$

$$z_{q[l],t} \geq 0, \forall q, \forall l \in \mathcal{H}, \forall t, \quad (19c)$$

## IV. CASE STUDIES

A case study on IEEE 24-bus system [21] is provided to demonstrate the effectiveness of the proposed methods. This test system contains 24 buses, 33 generators and 38 lines, which also has wind power as renewable resources. The base case has a total demand from 1,195 MW to a peak of 2,116 MW. Additional deviation ranging from [-20%, 20%] is applied to the base value. The mathematical model-based data generation is operated in Python using Pyomo. Regarding post-contingency frequency limits, RoCoF must be higher than -0.5Hz/s to avoid the tripping of RoCoF-sensitive protection relays, and the optimality gap is set to 0.1%. The PSS/E software is used for time domain simulation and labeling process. Full-scale models with detailed generator dynamics are implemented for more realistic data.

### A. Predictor Training

For the DNN layers, the number of neurons is set 10 for each layer. ReLU is used as the activation function. A total of 8,161 samples were collected based on strategies proposed in previous section. The entire dataset is divided into two subsets: 6,529 samples (80%) for training and 1,632 samples (20%) for validation. Fig. 2 presents the evolution of MSE losses on the training and validation sets over the training process of the proposed DNN model. Mean squared error (MSE) decreases as the number of epochs increases.

TABLE I shows the metrics being used to demonstrate the validation accuracy of the RoCoF predictor. $R^2$ of the predictor is 0.9828, indicating that the regression prediction fits the data well. The median absolute error (MED-E) and mean absolute error (MEA-E) are 0.0065 and 0.0117, respectively.

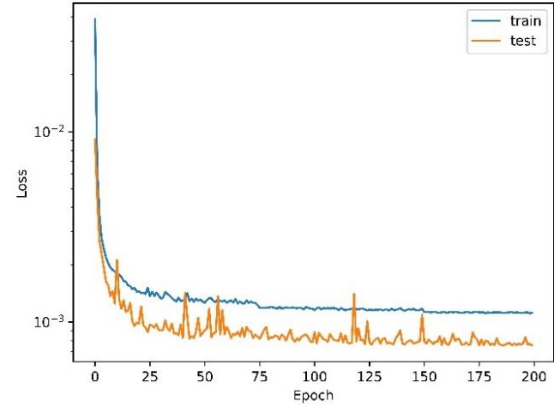

Figure. 2. MSE losses versus the number of epochs.

TABLE I
VALIDATION ACCURACY OF THE DNN-BASED ROCOF PREDICTOR

| Metrics | MED-E | MEA-E | $R^2$ |
|---|---|---|---|
| Prediction | 0.0065 | 0.0117 | 0.9828 |

### B. Simulation Results

The total scheduling horizon is 24 hours for four days. Hours 9-12 are selected to be the time instance where frequency related constraints are applied to secure system stability against generator contingency considering high penetration level of intermittent wind generation. The test case has a demand ranging from 1,604 MW to a peak of 1,916 MW. $RoCoF_{lim}$ is set to -0.5 Hz/s in this work.

TABLE II lists the simulation results of the proposed SLNN-RCUC model and other benchmark models. It can be observed that DNN-RCUC and the proposed SLNN-RCUC have the highest operational cost among all models; the extra cost results from the efforts in securing the RoCoF stability at each node. For RLNN-RCUC case, the total cost is same as T-SCUC model where no RoCoF related constraints are considered, implying that the approximation error due to ReLU linearization leads to RoCoF related constraints being non-binding.

TABLE II
COMPARISON OF DIFFERENT MODELS

| Model | Total Cost [$] | Computational Time [s] | Highest RoCoF [Hz/s] |
|---|---|---|---|
| T-SCUC | 1,486,556 | 13.58 | 0.8053 |
| ERC-SCUC | 1,494,431 | 20.24 | 0.6145 |
| LRC-SCUC | 1,615,135 | 35.25 | 0.5634 |
| DNN-RCUC | 1,641,967 | 368.58 | 0.4985 |
| RLNN-RCUC | 1,486,556 | 14.24 | 0.8053 |
| SLNN-RCUC | 1,641,953 | 66.78 | 0.4997 |

After inclusion of the complete ReLU, the resulting computational time of DNN-RCUC is increased to 368.58s. Incorporation of complete ReLU formulations introduces multiple binary variables into SCUC formulations, subsequently increases the computational time. With active ReLU linearization algorithm, the total computational time is reduced to 66.78 s. Besides the RoCoF related constraints remain binding in the case of SLNN-RCUC while it fails to bind in the case of RLNN-RCUC case.

Additionally, we run the dynamic simulation of G-1 contingency for all 5 models at hour 9 when the system netload is the lowest. As shown in TABLE III and Fig. 3, the proposed SLNN-RCUC can maintain RoCoF within safe range following contingency of generator loss. It should be noted that the ERC-SCUC and LRC-SCUC models cannot ensure system RoCoF security under same situation. Similarly, RLNN-RCUC model fails to secure RoCoF stability due to high approximation error.

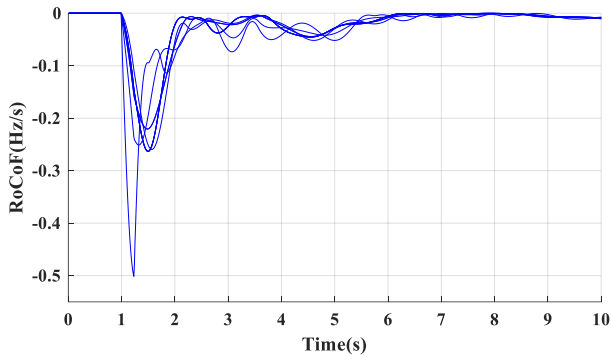

Fig. 3. RoCoF evolution of SLNN-RCUC model.

TABLE III indicates the difference in the total number of generators committed between the two runs. For hour 10, additional synchronous machines are committed to cover the loss of largest generation for SLNN-RCUC model, which accordingly increases the operation cost. While the total committed generator decreases at hour 11. Results suggest that, although RoCoF related constraints are incorporated in the SCUC model, the amount of committed generators only changes slightly.

TABLE III
COMPARISON OF THE TOTAL NUMBER OF DISPATCHED GENERATORS

| Hour | 8 | 9 | 10 | 11 | 12 | 13 |
|---|---|---|---|---|---|---|
| T-SCUC | 15 | 15 | 15 | 15 | 15 | 15 |
| SLNN-RCUC | 15 | 15 | 16 | 13 | 15 | 14 |

## V. CONCLUSIONS

In this paper, we proposed an SLNN-RCUC model to secure system RoCoF stability while maintain the computational efficiency of RCUC model. By incorporating DNN into T-SCUC model, system stability could be secured with less conservativeness. However, this process would significantly increase the computational burden of DNN-RCUC model. A node-of-interest active neuron selecting method is proposed to improve the computational efficiency of DNN based RCUC algorithm.

The simulation results on the IEEE 24-bus system indicate that the proposed SLNN-RCUC can improve power system inertial responses without much increased computational burden. Our proposed node-of-interest active selecting method has been proved to significantly reduce the computational time comparing to DNN-RCUC. And such algorithm can also reduce the approximation error of ReLU linearization, maintaining the performance of the proposed method.